%% file: imaging_chapter.tex
\documentclass[a4paper,11pt]{article}
\usepackage{aaskaiid}
\usepackage{xcolor}
\usepackage{orcidlink}
\title{Imaging the 21-cm Signal from the Cosmic Dawn \& Epoch of Reionization and the Connection with the Global Signal}
\ShortTitle{CD/EoR Imaging}

\author[1,2]{Satadru Bag\orcidlink{0000-0003-0141-606X}}
\ShortName{Bag et al.} 
\author[3,4]{Michele Bianco\orcidlink{0000-0002-6766-0017}}
\author[5]{Samir Choudhuri\orcidlink{0000-0002-2338-935X}}
\author[6]{Saswata Dasgupta\orcidlink{0000-0001-6461-769X}}
\author[7]{Abhirup Datta\orcidlink{0000-0002-5333-1095}}
\author[8]{Kanan K. Datta\orcidlink{0000-0002-2238-5146}}
\author[9,10]{Ivelin Georgiev\orcidlink{0000-0002-1950-5039}}
\author[9,11]{Sambit K. Giri\orcidlink{0000-0002-2560-536X}}
\author[12]{Ian Hothi\orcidlink{0000-0003-3356-5617}}
\author[13]{Akanksha Kapahtia\orcidlink{0000-0003-0348-0065}}
\author[7]{Suman Majumdar\orcidlink{0000-0001-5948-6920}}
\author[10]{Garrelt Mellema\orcidlink{0000-0002-2512-6748}}
\author[8]{Arnab Mishra\orcidlink{0009-0000-7477-682X}}
\author[7]{Samit Kumar Pal\orcidlink{0000-0002-2271-4165}}
\author[14]{Feiyu Zhao\orcidlink{0000-0003-1629-6021}}

\affiliation[1]{Max-Planck-Institut für Astrophysik, Garching D-85748, Germany}
\affiliation[2]{Technical University of Munich, TUM School of Natural Sciences, Physics Department, James-Franck-Straße 1, 85748 Garching, Germany}
\affiliation[3]{Laboratoire d'Astrophysique, École Polytechnique Fédérale de Lausanne (EPFL), Observatoire de Sauverny, Chemin Pegasi 51, CH-1290 Versoix, Switzerland}
\affiliation[4]{Institute for Particle Physics and Astrophysics, ETH Zurich, Wolfgang-Pauli-Str 27, CH-8093 Zurich, Switzerland}
\affiliation[5]{Centre for Strings, Gravitation and Cosmology, Department of Physics, Indian Institute of Technology Madras, Chennai 600036, India}
\affiliation[6]{Institute of Astronomy \& Kavli Institute for Cosmology, University of Cambridge, Cambridge, UK}
\affiliation[7]{Department of Astronomy, Astrophysics and Space Engineering, Indian Institute of Technology Indore, Indore 453552, India}
\affiliation[8]{Department of Physics, Jadavpur University, Kolkata, India}
\affiliation[9]{Department of Astronomy \& Oskar Klein Centre, Stockholm Universiy, AlbaNova Universiy Centre, Stockholm, Sweden}
\affiliation[10]{ARCO (Astrophysics Research Center), Department of Natural Sciences, The Open University of Israel, 1 University Road, PO Box 808, Ra’anana 4353701, Israel}
\affiliation[11]{Van Swinderen Institute for Particle Physics and Gravity, University of Groningen, Nijenborgh 3, 9747 AG Groningen, The Netherlands}
\affiliation[12]{Laboratoire de Physique de l'ENS, ENS, Université PSL, CNRS, Sorbonne Université, Université Paris Cité, 75005, Paris, France}
\affiliation[13]{Max-Planck-Institut für Astronomie, Königstuhl 17, D-69117 Heidelberg, Germany}
\affiliation[14]{Shanghai Astronomical Observatory, Chinese Academy of Sciences, 80 Nandan Road, Shanghai 200030, China}
\emailAdd{garrelt.mellema@astro.su.se}

\abstract{The original baseline design for SKA-Low was motivated by the ability to produce tomographic images of the redshifted 21-cm signal, thus allowing the research field to move beyond the simple statistic of the power spectrum. In this chapter we review the imaging capabilities of SKA-Low, the wide variety of methods proposed for quantatively analysing image data, as well as the connection with the global 21-cm signal.}

\begin{document}
\maketitle
\tableofcontents

\section{Introduction}

The original baseline design for SKA-Low \citep{BLD_document} envisaged an interferometer which would completely transform the study of the 21-cm signal from the Cosmic Dawn (CD) and Epoch of Reionization (EoR) by providing sufficient signal-to-noise ratio to enable imaging the signal at scales of a few arcminutes. This was revolutionary as none of the interferometers operational or planned at that time had images as its science goal; they rather focused on a detection of the statistical 21-cm power spectrum. Producing images of an intergalactic medium consisting of ionized regions embedded in still neutral areas will not only make the reionization process more visible and real for a wider science and lay audience, but also enable answering a range of questions which the power spectrum is unable to address.

The system currently under construction, with a first goal defined as AA* and a final goal as AA4, retains this imaging capability. The first results from AA* will focus on a statistical characterization of the signal in the form of power spectra. Given that none of the precursors has yet achieved a detection, this may be the first detection of the redshifted 21-cm signal. However, subsequent studies with SKA-Low will undoubtedly focus on extracting images, or rather image cubes, since the broad frequency coverage of SKA-Low allows tomographic imaging of the cosmological 21-cm signal.

The SKA Science Book from 2015 contained two chapters which specifically focused on tomographically imaging the 21-cm CD / EoR signal \citep{2015aska.confE..10M, 2015aska.confE..15W}. Those chapters still provide a good description of the basic concepts. The current chapter provides an update, considering the specific imaging capabilities of SKA-Low AA* and AA4 (Section 2). Furthermore, the past decade has seen major progress in the development of techniques to identify structures in noisy images, which we will summarize in Section 3. Section 4 will provide an overview of the large body of work describing different methods to quantitatively analyze tomographic image data. These quantitative analysis methods can be used in inference studies, and the chapter on inference \citep{Acharya01.2026.SKA} describes some examples of this. However, it is not necessary to identify structures in order to quantitatively analyze tomographic image data. Section 5 describes methods that can produce a quantitative characterization of the data without relying on the identification of structures. A large advantage of imaging data is that it is associated with a specific part of the sky and thus allows synergistic studies with other types of observables for that patch of sky. Section 6 provides an overview of what synergetic research is enabled by imaging data, which is described in more detail in the chapter on synergetic observations \citep{Chakraborty01.2026.SKA}. A special case is the connection with the global 21-cm signal, sometimes called the sky-averaged signal or the monopole. Section 7 describes how imaging can be used to derive the 21-cm global signal, as well as how short SKA baselines could be used to measure the global signal.

\begin{figure*}
    \centering
    \includegraphics[width=1.0\linewidth]{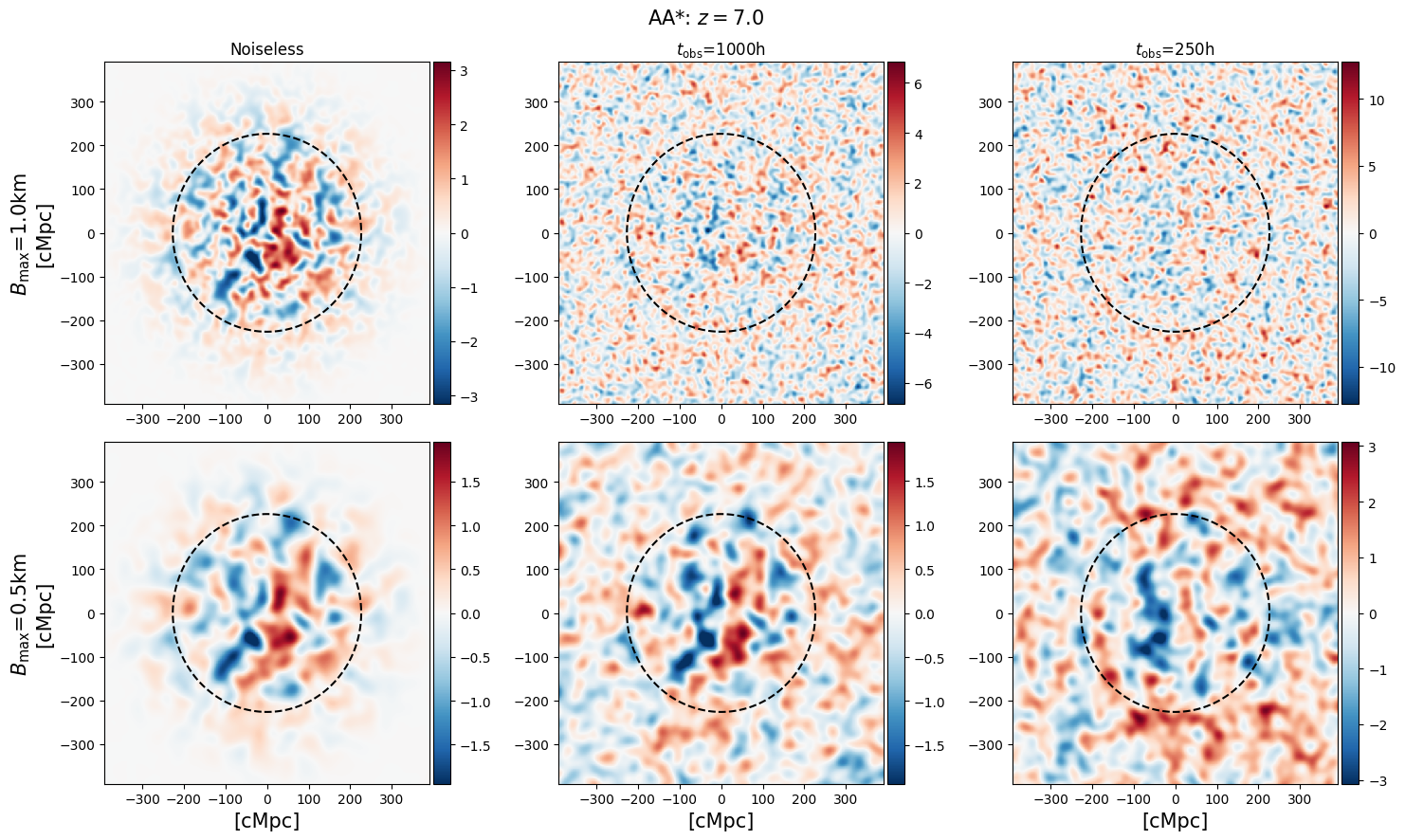}
    \caption{
    Sample 21-cm images for the AA* configuration of SKA-Low for a redshift $z = 7$ ($\sim$60\% ionized universe in the chosen model). The top row shows images using all baselines up to 1~km, the bottom row only uses baselines up to 500~m. The left column shows the noiseless image (including the primary beam), the middle column the image for a nominal integration time of 1000 hours, and right column the image for an integration time of 250 hours.
    }
 \label{fig:aastar_images}
\end{figure*}

\section{Imaging capabilities of AA* and AA4}

Figure \ref{fig:aastar_images} shows simulated images of the 21-cm signal with the AA* roll-out phase of SKA-Low. The three columns show the noiseless image, an image for an integration time of 1000 hours, and one with an integration time of 250 hours. The top row uses baselines up to 1 km (the full core), whereas the bottom row has half the resolution by only retaining baselines up to 500 m. The chosen redshift is $z=7$ ($\nu=177.6$~MHz). The bandwidth for these images is matched to their angular resolution and thus 0.92 and 1.85~MHz respectively. All images include the effect of the primary (station) beam which is assumed to have a Gaussian shape. The extend of its FWHM is indicated by a black dashed circle. The lower integration time of 250 hours can on the one hand be considered to give an indication of the possibilities of early imaging. However, as most interferometric images struggle to reach their theoretical noise level, they could also be seen as imperfectly calibrated images of 1000 hours of integration. These images do not contain any residual foreground effects and were produced with the {\sc tools21cm} package \citep{giri2020tools21cm}.

We can conclude that the high resolution ($B_\mathrm{max}=1$~km or $\sim 6^\prime$) images are heavily noise-dominated and cannot be used to extract any features. Reducing the resolution to $12^\prime$ yields a much better representation of the underlying structures, at least at the lower noise level. At the higher noise level, the features are clearly affected by noise and the analysis should be performed keeping this in mind.

\begin{figure*}
    \centering
    \includegraphics[width=1.0\linewidth]{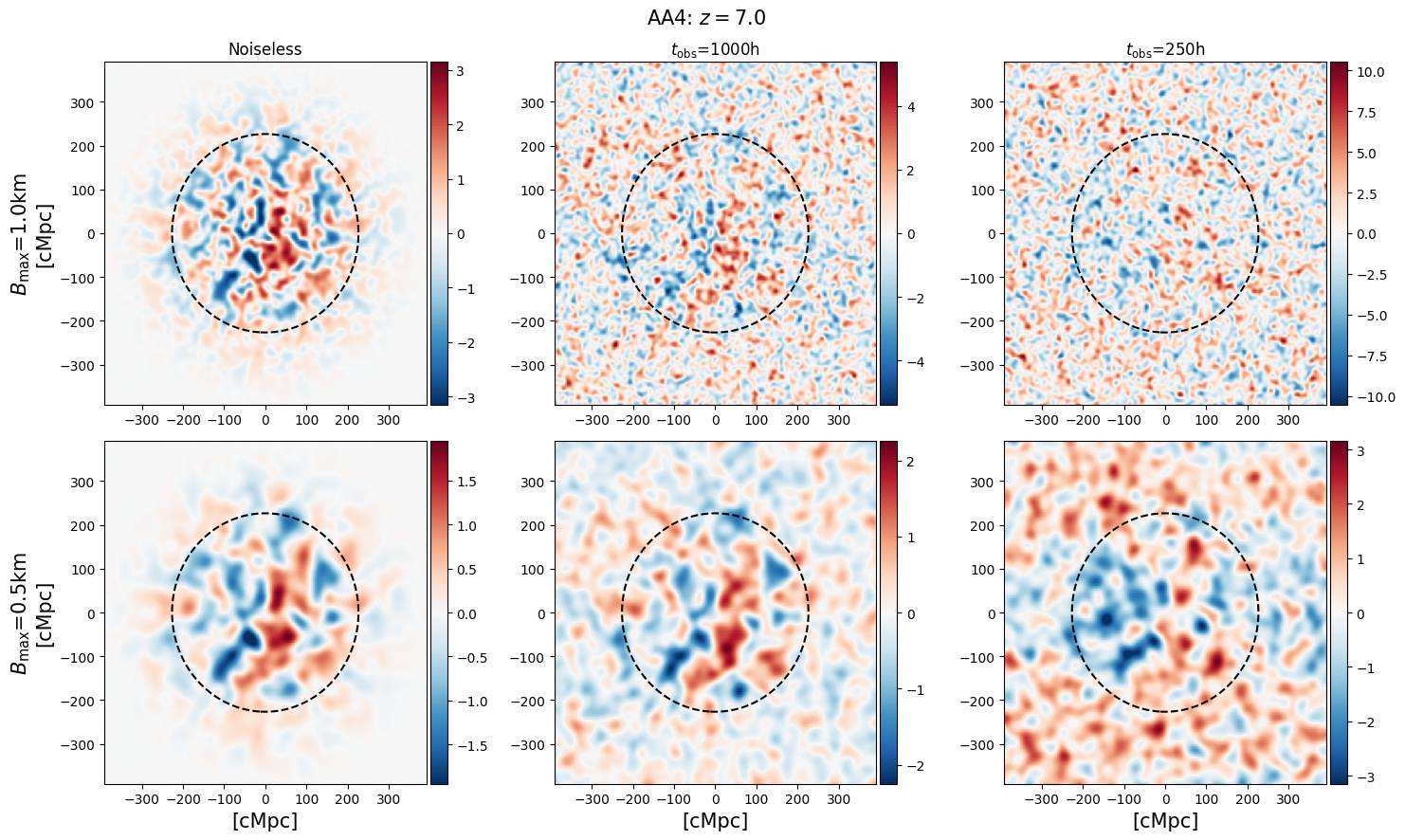}
    \caption{
    As Fig.~\ref{fig:aastar_images} but for the Baseline Design AA4.
    }
    \label{fig:aa4_images}
\end{figure*}

Figure~\ref{fig:aa4_images} shows the same images but now using the full Baseline Design, AA4. The additional sensitivity produces improvements for all cases. For example, the higher resolution, low noise case (middle panel top row) shows better correlation with the noiseless image. We can therefore conclude that the AA4 version of SKA-Low will have superior imaging to the AA* one.

\begin{figure*}
    \centering
    \includegraphics[width=\linewidth]{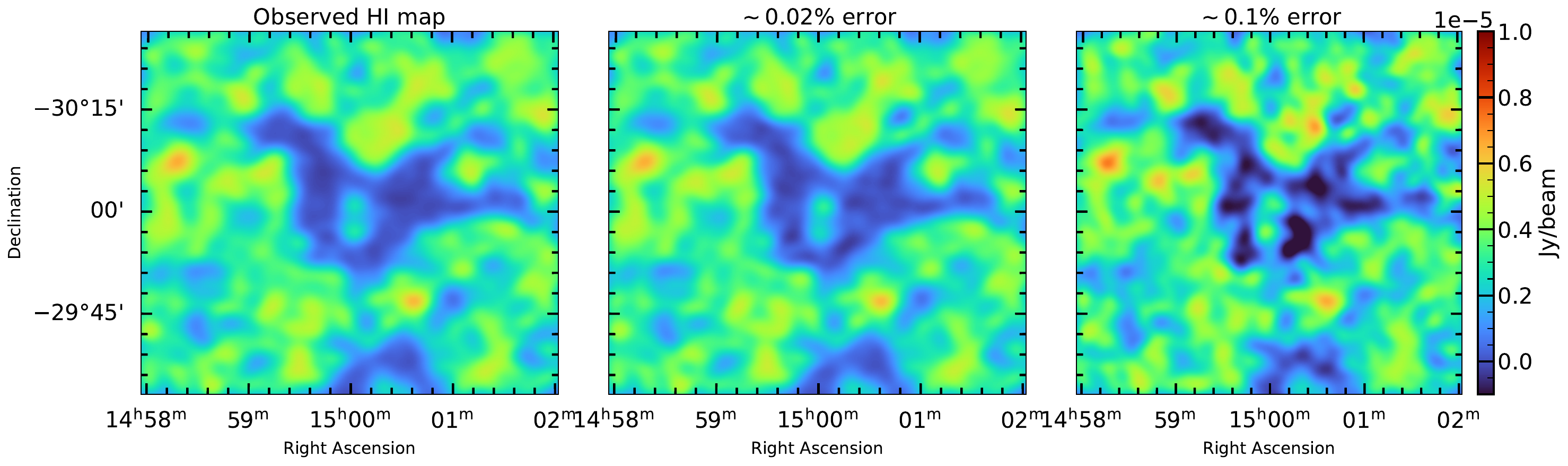}
    \caption{Impact of residual post calibration errors on the observed 21-cm maps via synthetic SKA-Low (AA4) observation at z = 7.76 ($\sim 40\%$ ionized universe in the chosen model) with angular resolution of $2.7' (7.1 \text{cMpc})$. Results from \citet{pal25}.}
    \label{fig:gain_err}
\end{figure*}

The above results are somewhat idealized as they did not consider the impact of residual foregrounds or calibration errors. Some insights on the impact of those observational effects can be obtained from the results of \citet{Dasgupta2023}. Those authors produced realistic SKA-Low 21-cm maps simulated via a synthetic radio observation as per the characteristics of the upcoming SKA-Low AA4 configuration. The synthetic maps were produced using the 21cmE2E pipeline \citep{2022Mazumder, 2024Pal}, based on OSKAR\footnote{\url{https://ska-telescope.gitlab.io/sim/oskar/}} software for SKA-Low, and CASA\footnote{\url{https://casa.nrao.edu/}}. They used different deconvolution algorithms (such as H\"ogbom and Multiscale) and various weighting schemes (natural and Briggs weighting) to investigate their effects on the morphological features of ionized regions. \citet{pal25} extended this analysis by introducing the effects of antenna-based gain calibration errors into the 21-cm observation maps. Figure~\ref{fig:gain_err} illustrates these effects, showing how varying levels of residual post-calibration errors affect the reconstruction of the 21-cm maps. This work highlights how calibration errors and residual foreground contamination artifacts can introduce artificial structures which can affect the identification of structures (see Section \ref{sec:structures}). 

These results highlight that although SKA-Low in principle can produce tomographic images of the 21-cm signal, it will not be trivial to produce data in which structures can be robustly identified. Both low noise levels and excellent calibration and foreground removal will be critical to achieve high image fidelity.

\section{Identifying structures in images}
\label{sec:structures}
As is evident from the results in the previous section, image cubes from SKA-Low can be expected to contain a fair level of noise and artifacts such as residual foreground structures or instrumental effects. Most of the techniques for analyzing tomographic image data are based on analyzing structures seen in the data, e.g.\ ionized regions or neutral islands. The first step is therefore to identify reliably such structures in noisy data. This section provides an overview of the different methods that have been proposed to achieve this. In the first subsection, we consider techniques that work on the image data, whereas the second subsection describes how certain types of structure can be identified directly in the visibility data.

\subsection{Structures from image data} \label{sec:structures_images}

The 21-cm signal image data contains patterns that can be exploited to identify structures. For example, the 21-cm signal is produced by the neutral areas in the IGM, which in absence of spin temperature fluctuations leads to a bimodal distribution of ionized areas with no signal and neutral areas with signal. In the neutral areas the signal will have a range of values due to the density fluctuations. The response of the interferometer may still assign some specific value to the areas without signal but that can be considered as a constant offset. It is this bimodal distribution of signal values which allows the separation of ionized and neutral regions.

The actual separation will be complicated by the presence of noise, residual foregrounds, and finite resolution. Several methods have been proposed to choose a threshold value to segment image data into ionized and neutral regions. \citet{KakiichiMajumdar_2017} simply used the mean value of the image data as the threshold. \citet{GiriMellema_2018a} proposed using the K-means clustering algorithm \citep[see][for a review]{kodinariya2013review} to separate ionized and neutral resolution elements. \citet{giri2018optimal} extended this approach adding the maximum deviation method to deal with the cases when the distribution is no longer clearly bimodal \citep[see also][]{Gazagnes+2021}. \citet{Dasgupta2023} instead used a gradient-descent method to find the local minima between the two peaks of the bimodal distribution. Yet another method was proposed by \cite{pal25} in which unsharp masking is used to enhance edges between neutral and ionized regions before applying a global thresholding method.

A more advanced method was proposed by \citet{giri2018optimal}. It not only considers the value of the resolution elements, but also their environment by implementing the Simple Linear Iterative Clustering (SLIC) method. This method calculates distances in the four dimensional space of position and 21-cm signal. The grouping of resolution elements according to the SLIC method leads to the Superpixel method for image segmentation. This method can reconstruct the bimodal distribution even in the presence of noise. It was further extended to employ cross-entropy in \citet{giri2025mapping} to classify these superpixels into ionized and neutral regions.

Another approach to segment images into ionized and neutral regions is to train machine learning methods to recognize meaningful patterns by training it on simulated noisy data. For example, neural networks with U-shaped architecture can learn a mapping between two image datasets \citep{ronneberger2015u}. These networks have been employed to learn how to construct binary maps of ionized/neutral regions from noisy 21-cm images \citep[e.g.,][]{Bianco2021deep}. This approach has been quite successful in identifying ionized/neutral regions in these images even in the presence of foreground residuals \citep[e.g.,][]{gagnon2021recovering, Bianco2024deep,kennedy2024machine}. We refer the interested readers to the chapter on machine learning methods \citep{Acharya02.2026.SKA} for more discussion.

The above studies focus on segmenting image data into neutral and ionized regions, assuming the bimodal distribution of signal values which can be expected when there are no substantial spin temperature fluctuations. During the Cosmic Dawn when the ionized regions are still very small, we expect such spin temperature fluctuations due to local differences in Lyman-$\alpha$ coupling and X-ray heating. Under these circumstances it may be much harder to segment the image data as spin temperature differences are expected to produce a much more continuous distribution of signal values. Given this it could of course be argued that segmentation of images into two different regimes is not really useful. Still, the problem of how to quantitatively analyse 21-cm image data in case of substantial spin temperature fluctuations has to date not been systematically studied, even though the quite large scales on which differences are likely to exist may allow the construction of very low resolution images of the Cosmic Dawn from SKA-Low observations \citep{2021MNRAS.506.3717R}.

\subsection{Structures from non-image data}

If it is known what kind of structure you are looking for, it is possible to look for its signature in the visibility data, without the need to construct images. This is the idea behind the so-called matched-filter method \citep{Datta_2007, Datta_2008, Datta_2012,Majumdar_2011, Majumdar_2012, mishra2024}. This method searches for isolated spherical ionized regions which are expected to exist around high-redshift luminous sources such as quasars and galaxy clusters \citep{Mortlock_2011, Ba_ados_2017, Wang_2018, Matsuoka_2019, Matsuoka_2019a, Wang_2021, witstok24}. As it is working directly on the visibilities it is very efficient and this method may be particularly useful if it is hard to produce reliable images. This approach has already been applied to existing data from the MWA \citep{2021MNRAS.507..772T} and HERA \citep{Chen+2025}.

\begin{figure}
\centering
\includegraphics[width=0.49\textwidth]{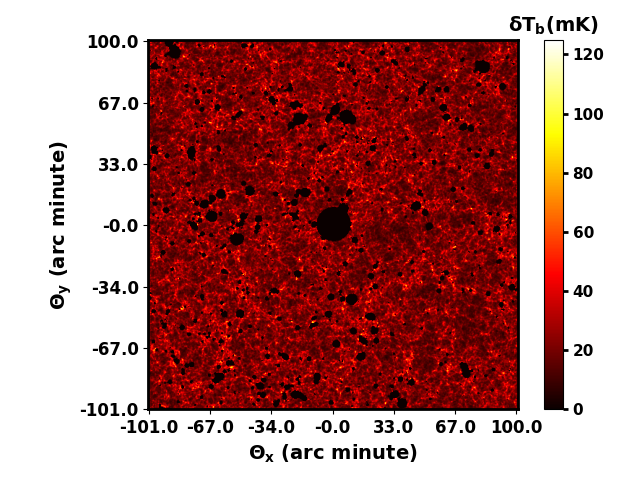}
\includegraphics[width=0.49\textwidth]{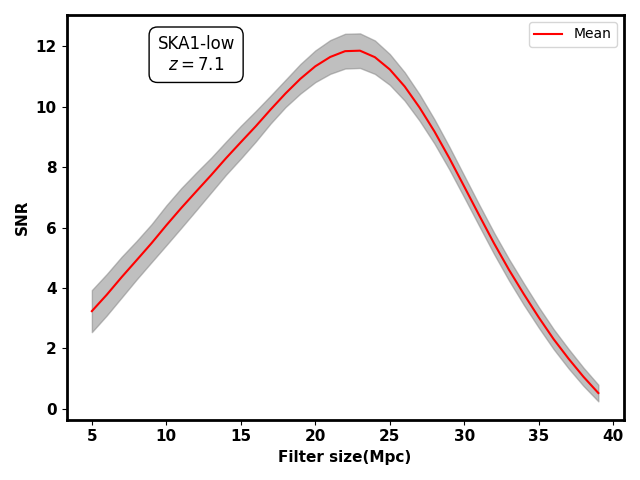}

\caption{Left panel: Differential brightness temperature map of the HI 21-cm signal around an ionized bubble of comoving size $\sim 24$ Mpc at $z=7.1$ corresponding to an observing frequency of $175$ MHz. The angular size of simulated maps is $\sim 3.3^{\circ}\times 3.3^{\circ}$ with $ \sim 23^{\prime\prime}$ resolution. Right panel: SNR of 20 independent noise realizations as a function of filter size for $100$ hours of the SKA-Low AA4 observations at redshift $7.1$. The shaded region shows $1\sigma$ uncertainty estimated from independent realizations. Both panels are taken from \citet{mishra2024} for reference.}
\label{fig:HII_mortlock}
\end{figure}
As an example, the left panel of figure~\ref{fig:HII_mortlock} shows a simulated HII bubble around a bright quasar at redshift $z=7.1$, similar to the one reported in \citet{Mortlock_2011}. The map is simulated using parameters consistent with a SKA-Low AA4 observation. The input bubble parameters used are: radius $R_b \sim 24 \,\mathrm{Mpc}$, centered in the field of view, and mean neutral hydrogen fraction $x_{\mathrm{HI}} = 0.88$. A $100$-hour observation with the $AA^{\star}$ and $AA4$ configurations yields matched-filter detections with ${\rm SNR} \approx 10.3$ and ${\rm SNR} \approx 11.8$ (right panel of Figure \ref{fig:HII_mortlock}), respectively for this HII bubble. The number of short baselines in the SKA-Low $AA^{\star}$ and $AA4$ configurations is very similar; however, $AA4$ includes significantly more long baselines. Consequently, the difference in the SNR between the two configurations is small, since the HI 21-cm signal contributes very little at large baselines. However, the long baselines play an important role in foreground mitigation, particularly for modelling and subtracting point sources.  Studies further show that the SNR scales with bubble size approximately as $\mathrm{SNR} \propto R_b^{1.7}$ for SKA-Low $AA^{\star}$ and $\mathrm{SNR} \propto R_b^{1.95}$ for SKA-Low $AA4$. This highlights that larger ionized regions can be detected more readily, and the scaling relation provides a useful guide for predicting detectability across a range of bubble sizes and observational setups. 

\begin{figure}
\centering
\includegraphics[width=0.49\textwidth]{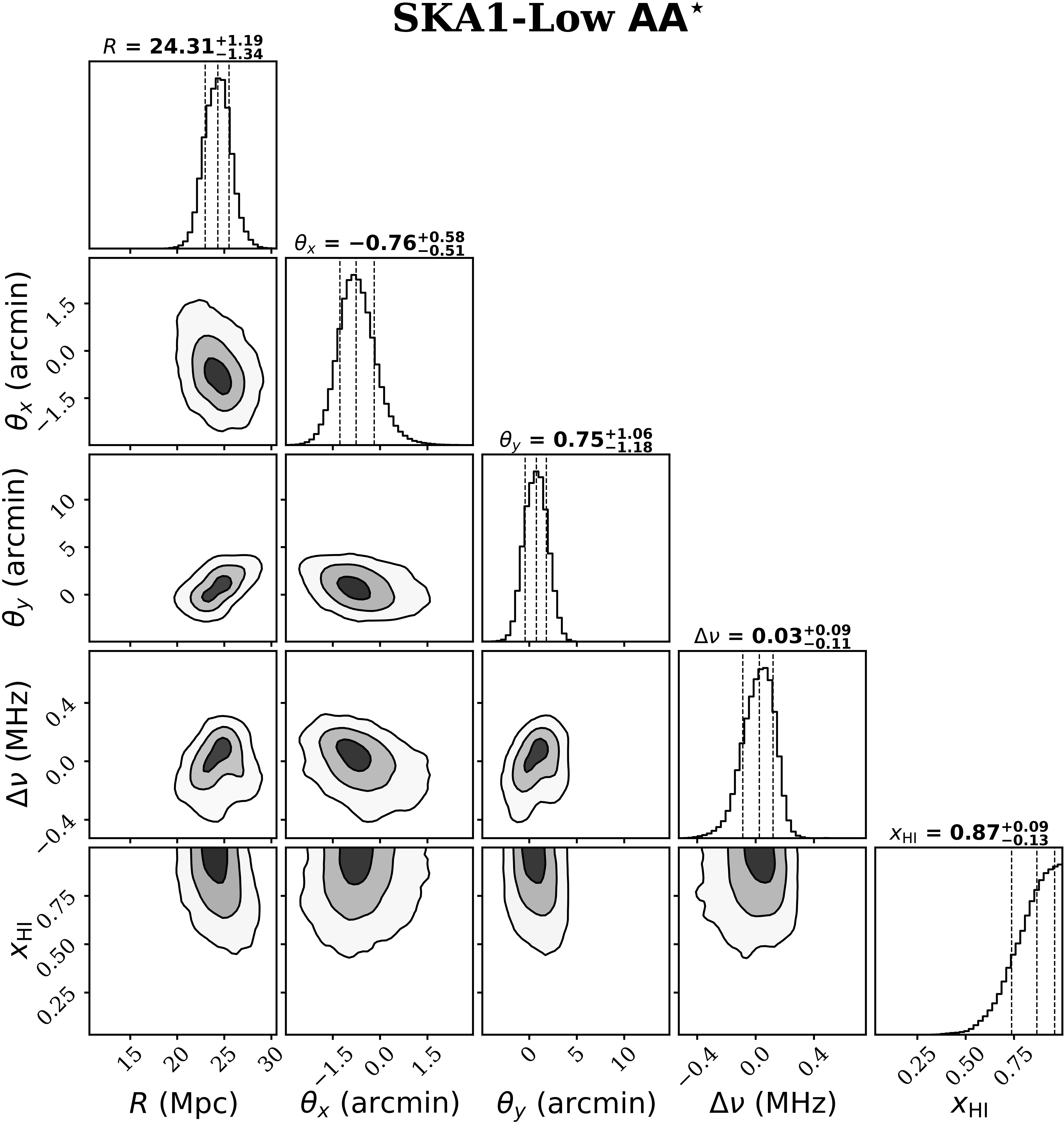}
\includegraphics[width=0.49\textwidth]{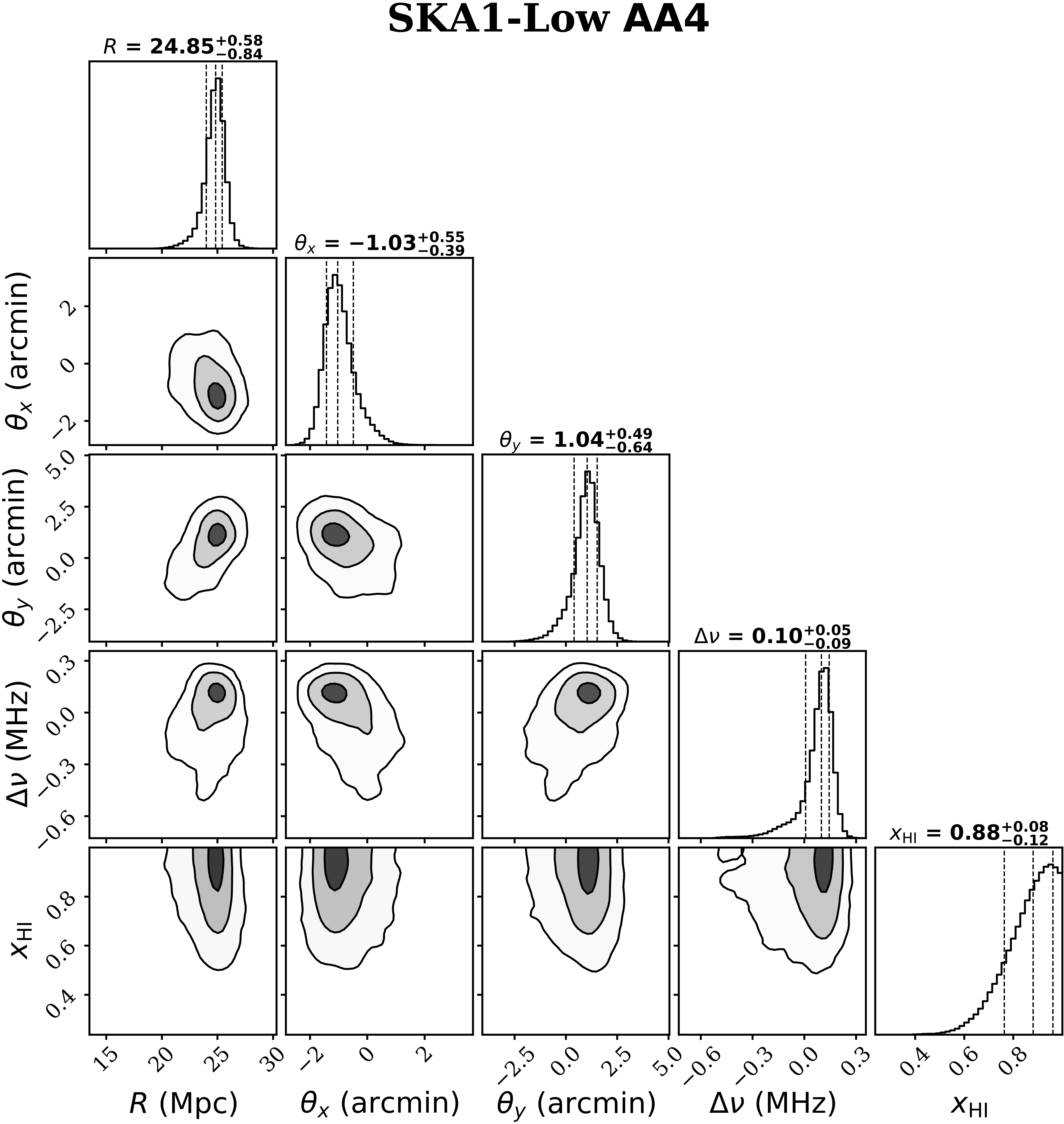}

\caption{Posterior probability distributions of ionized bubble parameters derived from Bayesian analysis of the matched-filter outputs. The left panel corresponds to SKA-Low $AA^{\star}$ and the right panel to SKA-Low $AA4$. The contours represent the $1\sigma$, $2\sigma$, and $3\sigma$ credible intervals, constraining bubble size, position, frequency offset, and neutral hydrogen fraction.}
\label{fig:corner_plot}
\end{figure}

A Bayesian parameter estimation framework \citep{Ghara_2020} has also been applied to constrain bubble properties such as size and location, as well as IGM parameters such as the neutral hydrogen fraction. The resulting posterior distributions are shown in Figure~\ref{fig:corner_plot}. These results demonstrate that a visibility-based matched filtering, combined with Bayesian inference, provides a robust framework for detecting and characterizing isolated ionized bubbles using SKA-Low $AA4$ and $AA^{\star}$ observations.

\section{Analysing structures identified in images}

Once structures have been identified, the goal is to quantitatively analyse them. As the structures characteristic for CD/EoR are expected to be complex in shape, and when considering tomographic data, are actually three-dimensional, their analysis is far from trivial. In fact, there exist many different ways to analyse them. Over the past 10 years a considerable amount of effort has been dedicated to exploring different kinds of quantitative methods for describing the morphology of regions. This section provides an overview. The most obvious quantitative description is one which captures the sizes of identified structures and the first subsection is dedicated to such size distributions. The mathematical theory for describing shapes in multiple dimensions is known as topology, and the second subsection describes how topological quantities can be used to characterize structures in image data. 

\subsection{Size distributions}
\label{sec:BSDs}
The size distribution and evolution of ionized bubbles during reionization are closely tied to the astrophysical processes driving the growth of ionization fronts and to the statistical properties of the ionization field \citep{FurlanettoMcQuinn_2006}. Bubble size distributions (BSDs) quantify the characteristic scales of these regions and provide valuable information about the timing, morphology, and topology of reionization. A range of methods has been developed to model BSDs \citep{furlanetto2004growth,DoussotSemelin_2022} and to estimate them from simulated or observed 21\,cm signal data \citep{GiriMellema_2018a,LinOh_2016}.

One of the earliest approaches is the friends-of-friends (FoF) algorithm \citep{IlievMellema_2006}, which identifies connected ionized pixels and groups them into contiguous regions, from which bubble sizes are derived. As this method uses connected regions, it is closely connected to the topological quantities discussed below in Section~\ref{sec:topological}. The typical size distribution produced by the FOF method does not resemble the other size distributions described next. As described in \citet{2016:Furlanetto-percolate}, after percolation the FOF size distribution is dominated by one large connected region, the percolation cluster. The volume fraction of this dominant cluster relative to the total ionized volume is called the largest cluster statistic (LCS) and will be discussed in more detail below.

The spherical average method (SPA) \citep{ZahnLidz_2007} derives a BSD by finding, for each ionized pixel, the largest sphere centered on it for which the average ionized fraction exceeds a chosen threshold. This method has been found to quite diffusive and has therefore become less popular. A more popular algorithm is the mean free path (MFP) method \citep{21cmFAST_2007,21cmFAST_2011}, implemented for example in \textsc{21CMFAST}, which employs a Monte Carlo procedure: random ionized pixels are selected, and the distance to the nearest neutral pixel along a random direction is recorded; the resulting histogram of distances traces the bubble size distribution. Another estimator for size distribution uses granulometry \citep{KakiichiMajumdar_2017}. In this method the ionization field is repeatedly sieved with spherical holes of increasing radius $R$ and the fraction of ionized pixels removed at each step is used to build a probability distribution of bubble sizes. \citet{LinOh_2016} introduced a three-dimensional extension of the watershed transform, which segments the ionization field into regions by identifying isodensity surfaces that trace bubble boundaries. 

Each of these four methods produce somewhat different size distributions for the same input data. However, the distributions are comparable in that they typically display a peak value and a roughly log-normal distribution of sizes. Comparisons of these methods applied to simulated maps without noise or foreground contamination have been presented in \citet{LinOh_2016}, while \citet{GiriMellema_2018a} extended the analysis to include observational effects.  

More recently, machine learning approaches have been explored; for example, \citet{Shimabukuro2022} trained artificial neural networks to recover MFP size distributions directly from the 21\,cm power spectrum, demonstrating that this direct recovery yields higher accuracy than inferring size distributions from model parameter fits. However, \citet{2025JCAP...06..003C} showed artificial neural networks can only recover size distributions from power spectra produced by the reionization code they were trained on. 

\subsection{Topological quantities}
\label{sec:topological}
In a 2D scalar field such as the 21-cm brightness temperature map, the excursion set is defined as regions above a certain chosen threshold. Its shape and connectivity captures both the morphology and topology of the underlying field. These excursion sets evolve during the EoR: ionized regions grow from isolated bubbles into a percolating, connected network, while neutral regions fragment over time \citep{2016:Furlanetto-percolate}. The Friends-of-Friends size distribution introduced in Section~\ref{sec:BSDs} is in fact the size distribution of the region defined by the excursion set. Real-space statistics from images retain both amplitude and phase information, making them ideal for probing the highly non-Gaussian 21-cm field. 

Perhaps the simplest topological quantity which can be defined is the Largest Cluster Statistic (LCS), already briefly mentioned above in Section~\ref{sec:BSDs}. This single number, defined as
\begin{equation}
{\rm LCS}=\frac{\rm volume ~of ~the ~largest ~ionized ~region}{\rm total ~volume ~of ~all ~the ~ionized ~regions}\;,
\end{equation}
has been shown to be able to distinguish extreme reionization models \citep{pathak2022distinguishing} and thus could be a simple but useful quantity to consider. A series of papers has further explored the potential of the LCS, also considering the practical challenges of measuring it in noisy and imperfectly calibrated image data \citep{Dasgupta2023, pal25}. For this and the other topology-related quantities described in this section, both noise, residual foreground and artifacts introduced by calibration errors can be especially damaging as they affect the connectivity. \cite{pal25} concludes that for a minimum of 2000\,hours of SKA-Low (AA4) integration and for a maximum antenna-based calibration error tolerance of $\sim 0.02\%$ after calibration, the reionization history can be recovered in a robust and relatively unbiased manner using LCS. 

Morphological and topological descriptors have been widely used to study neutral and ionized regions in simulations and mock 21-cm images during reionization.  One of the earliest morphological descriptors used for EoR fields is the \textit{Scalar Minkowski Functionals (SMFs)}. In $n$ dimensions, there are $n+1$ SMFs. In 2D, they are contour length, area fraction, and genus (connected regions minus holes) of the excursion set. The genus has been used to study EoR phases \citep{2008:Lee} and distinguish reionization scenarios in noisy mock 21-cm images \citep{2014:Hong,2015:Wang}. In 3D—relevant for SKA-low observations—SMFs are volume, surface area, integrated mean curvature, and Euler characteristic. The latter is a single number that expresses the topology of a field by counting connected components, tunnels, and cavities. These have been used to study reionization morphology \citep{2006:Gleser, 2011MNRAS.413.1353F, YoshiuraShimabukuro2017,2019:chen}. It has been shown that when used in conjunction with the 21-cm power spectrum, the 3D SMFs give a 30\% improvement in constraining reionization model parameters than when using the power spectrum alone \citep{2024:Diao}.

Another interesting approach focuses on the geometry of individual ionized or neutral regions, including their morphological and topological properties. In $n$ dimensions, $n$ {\em Shapefinders} (SFs) are defined as ratios of $(n+1)$ SMFs, with each SF having the dimension of length and characterizing the spatial extension of a region bounded by a closed surface \citep{Sahni:1998cr}. SFs can be further combined to quantify the morphology of individual regions, e.g., their {\em planarity} and {\em filamentarity} in 3D.  Using the shape diagnostic tool \texttt{SURFGEN2} \citep{Sheth:2002rf,Sheth:2005ys}, which implements a Marching Cubes 33 triangulation scheme to model the surfaces in 3D, \citet{Bag18} showed that large ionized regions maintain a nearly constant cross-section at the percolation transition, while their lengths grow sharply, rendering them highly filamentary. The morphological properties of these regions, such as their characteristic cross-sections and the distributions of planarity and filamentarity at the onset of percolation, can be used to distinguish drastically contrasting reionization scenarios \citep{pathak2022distinguishing}.
From the excursion set approach, the morphology of HI overdense and underdense regions similarly shows that, as reionization progresses, the cross-sections of large overdense regions become increasingly uniform \citep{bag19}. 

Minkowski tensors (MTs) are generalizations of the SMFs, of which the \textit{Contour Minkowski Tensor (CMT)} is the only linearly independent, translation-invariant 2D MT that provides anisotropy information beyond SMFs \citep{2017:pravabati}. Since the eigenvalues of CMT encode both  scale and shape information, it was shown that it could be used to define a characteristic ionized bubble size, without assuming spherical symmetry \citep{2018:Kapahtia}. By mapping the morphology of the ionization, density, and spin temperature fields to that of the observable 21-cm brightness temperature \cite{2019:Kapahtia} qualitatively showed that the CMT could serve as a promising tool for discerning various sources of non-Gaussianity in the 21-cm field. 

\textit{Betti numbers} ($\beta_k$) are topological invariants that count $k$-dimensional holes in the excursion set. In 3D, $\beta_0$ counts isolated regions, $\beta_1$ tunnels, and $\beta_2$ cavities \citep{2019:Pratyush}. They are related to the Euler characteristic $\chi$ through
\begin{equation}
    \chi=\beta_0-\beta_1+\beta_2\,,
\end{equation}
and thus carry more information than this SMF. They have been used to trace reionization stages and percolation in mock 21-cm cubes \citep{2021:Giri}, showing that they evolve in a characteristic and parameterizable way, from which meaningful topological insights can be obtained. It has also been shown that when combined with the CMT, can constrain EoR model parameters from a smoothed noisy image at a single frequency slice \citep{2021:Kapahtia}. 

Tracking Betti numbers as a function of a parameter (filtration) such as the field threshold, allows for \textit{persistent homology}, which measures the birth and death of features as the filtration is varied. The richer topological information reveals distinct post-overlap reionization stages \citep{ElbersvandeWeygaert_2018} and shows prospects for distinguishing astrophysical models of EoR with 1000h of HERA data \citep{ElbersvandeWeygaert_2023}.

\section{Analysing images without identifying structures}

Instead of focusing on structures or shapes identified in image data, it is of course also possible to analyse the image data directly, that is, to subject it to some mathematical operation which should capture some of its essential properties. Actually, deriving the power spectrum is in essence such a method. However, as described elsewhere, the power spectrum is not sensitive to the non-Gaussian nature of the 21-cm fields. In this section we describe some methods proposed to process image data in order to capture its non-Gaussian character. 

\subsection{Scattering transform}

Scattering transforms have emerged as powerful summary statistics for characterizing non-Gaussian fields \citep{2011:Mallat, 2012:BrunaMallat}. Inspired by convolutional neural networks (CNNs), the scattering transforms rely on successive wavelet transforms, that separate a given field on its different oriented scales, as well as non-linear operations such as modulus, which allow to efficiently characterize the interactions between these different scales. Several generations of Scattering Transform statistics have been introduced : the initial Wavelet Scattering Transform (WST)~\citep{2019:AllysLevrierZhang, 2020:ChengTingMenard}, Wavelet Phase Harmonics (WPH)~\citep{2020:AllysMarchandCardoso}, and the Scattering Covariance~\citep{2023:ChengMorelAllys}. While the WST is highly compact set of summary statistics for characterization and parameter inference, the WPH and Scattering Covariances allow for additional application as generative modeling and component separation~\citep[see, for instance][]{2024A&A...681A...1A}

To get a sense of scattering transform statistics, let us consider the WST. Starting from an input field $I(x)$, the first layer of WST is built from the convolution with a set of wavelets parametrized by scale $j$ and orientation $\theta$:
\begin{equation}
S^1_{j, \theta} = \langle |I \star \psi_{j, \theta}| \rangle,
\end{equation}
where convolution is denoted by $\star$ and spatial average by $\langle \cdot \rangle$. These first-order coefficients $S^1_{j,\theta}$ measure the amplitude of fluctuations at various oriented scales, resembling a scale-resolved power spectrum, but allowing to characterize sparsity and local structures~\citep{2020:ChengTingMenard}.
To characterize interactions between scales, a second convolution with the set of wavelet if performed:
\begin{equation}
S^2_{j_1, \theta_1, j_2, \theta_2} = \langle ||I \star \psi_{j_1, \theta_1}| \star \psi_{j_2, \theta_2}| \rangle.
\end{equation}
These second-order scattering coefficients directly characterize the interaction between two scales. They capture non-Gaussian, higher-order correlations in the data, thereby circumventing the computational issues linked to conventional N-point statistics \citep{2019:AllysLevrierZhang,2023:ChengMorelAllys}.

Unlike the bispectrum, which characterises the interaction between individual Fourier modes, the scattering transforms rely on wavelets. These are bandpass filters that span an ensemble of Fourier modes. This broader mode coverage results in significantly lower variance for the estimated statistics, thereby improving robustness and stability~\citep{2023:ChengMorelAllys}. 

Scattering Transforms have already been applied to EoR simulations for parameter constraints~\citep{2022:GreigTingKaurov, HothiAllys_2024, Prelogovic24,2024:ZhaoMaoZuo,2025:ShimabukuroXuShao}, as well as to build generative model of EoR lightcones~\citep{2025:HothiAllysSemelin}. 

\subsection{Neural networks}
Beyond summary statistics, machine learning techniques, particularly Convolutional Neural Networks (CNNs), offer a powerful paradigm for "end-to-end" parameter inference. This approach aims to directly map 21-cm images (or image cubes) to the underlying astrophysical parameters, bypassing human-defined summary statistics such as the power spectrum or size distributions. The primary motivation is that NNs are exceptionally adept at identifying the complex, non-linear, and non-Gaussian patterns in the 21-cm data that other statistics may miss. Several studies \citep[e.g.,][]{Gillet:2019,Hassan:2019} have demonstrated the feasibility of this approach, showing that CNNs can recover key reionization physics from 21-cm images. Subsequent research has focused on improving robustness against realistic observational challenges such as instrumental noise and foreground contamination \citep{Prelogovic:2022,Zhao:2022a}. For detailed discussion, we refer the interested readers to the chapter on machine learning techniques \citep{Acharya02.2026.SKA}.

\subsection{Other analysis methods}
In addition to the methods already discussed, several other statistics have been proposed to extract non-Gaussian information from 21-cm images. The local variance, for example, is a one-point statistic that leverages sample variance along the frequency direction of an image cube to trace the reionization history and ionization morphology \citep{Gorce2021localvariance}. Marked statistics, such as the marked power spectrum, re-weight the field based on local density, allowing for enhanced sensitivity to information in low- or high-density regions that the standard power spectrum may not capture well \citep{Kamran2025marked}. The evolving morphology of the neutral hydrogen distribution can also be characterized using generalized fractal dimensions, which quantify the complex, multifractal nature of the ionized regions and can be used to constrain the reionization history \citep{Bandyopadhyay2017fractal}.

\section{Combining image data with other observables}
Image data contains positional information and so allows the 21-cm data to be combined with other observables. This section addresses some combinations of image data with data from different telescopes.

\subsection{Cross-correlation of images with other observables}

The most obvious path to combine the tomographic image data with other observables is to cross-correlate them. This has the added benefit that as calibration errors, foreground residuals and general systematics will be uncorrelated between the two data sets, the correlation can be studied even if the 21-cm image data is severely affected by such effects. As the synergy chapter \citep{Chakraborty01.2026.SKA} contains an extensive description of the different kinds of cross-correlations that have been explored, we will not repeat them here but refer to that chapter. However, the next section does discuss one particular cross-correlation study which is not contained in that chapter.

\subsection{Ly-$\alpha$ forest - IGM tomography}

The Lyman-$\alpha$ forest, observed as absorption features in the spectra of distant quasars, is a cornerstone probe of the IGM \citep[e.g.,][ for a review]{2023ARA&A..61..373F}. It provides 1D tomography of the (mostly ionized) IGM along individual lines-of-sight. This technique is most powerful at $z \lesssim 6$; at higher redshifts, the rapidly increasing IGM neutral fraction renders the forest completely saturated (the "Gunn-Peterson trough") \citep[e.g.,][]{Becker2015evidence}, limiting its utility deep in the EoR. A complementary technique, photometric IGM tomography, aims to move beyond 1D skewers by using deep narrow-band (NB) imaging of numerous faint background galaxies to reconstruct a 2D or 3D map of the Ly-$\alpha$ forest transmission \citep[e.g.,][]{Kakiichi2022photometric}. This mapping approach is particularly promising for identifying the final, large-scale neutral islands at the very end of reionization ($z \sim 5.7$) \citep{giri2025mapping}.

This is precisely where 21-cm tomography with SKA-Low becomes the dominant probe, as it is designed to directly image the 3D distribution of neutral gas within the EoR. A key synergy is the cross-correlation of SKA-Low 21-cm maps with Lyman-$\alpha$ forest data, which can be done using the reconstructed 2D/3D maps from Lyman-$\alpha$ forest transmissions. As demonstrated by \citet{giri2025mapping}, these reconstructed Lyman-$\alpha$ transmission maps (which have high values in ionized regions) and the SKA-Low 21-cm maps (which have high values in neutral regions) are expected to be strongly anti-correlated. A detection of this anti-correlation would provide a "smoking gun" confirmation of the cosmological 21-cm signal, as the foregrounds and systematics for radio and optical/NB surveys are entirely independent.

\subsection{Connecting individual ionized bubbles to the ionizing sources: a multiwavelength synergy with SKA}

Apart from cross-correlation another approach is multi-frequency studies of particular areas. If the 21-cm image data from SKA-Low identifies certain ionized areas, follow-up observations (or archival observations) in optical and near-infrared, of the same patch of the sky, will reveal the galaxy population there. As it is these galaxies which produced the photons that made the ionized region, such a joint study can reveal how galaxies reionized the Universe, and for example give constraints on the escape fraction of ionizing photons.

\citet{zackrisson_2020, mishra2024, 2015aska.confE..10M, Lu+2024, GiriMellema_2018a, Chen+2025, Mason+2020, Gazagnes+2021, Ghara+2018, Ghara_2020} and others suggest that SKA-Low will be able to detect ionized regions of size $\geq5-10$ cMpc, corresponding to angular scales exceeding $5$ arcminutes at redshifts $z=7-10$. This resolution threshold enables the detection of ionized bubbles with volumes $\geq 1000\,({\rm cMpc})^3$. Several of the above studies independently established, via various reionization simulations, that there is a strong correlation between the volume of the ionized regions and the number of escaped ionizing photons (mostly UV) from the sources of ionization residing inside those bubbles. 

Further \citet{zackrisson_2020} suggest that using JWST, Euclid, ELT, Nancy Grace Roman Space Telescope, etc., one can, in principle, detect the rest-frame UV emission from star-forming galaxies within ionized bubbles already detected via SKA-Low. However, the number of detectable bubble galaxies, either via photometry or spectroscopy, will depend on the time-integrated escape fraction of ionizing photons and ionizing photon production efficiency. In the case of a galaxy-dominated reionization model, an escape fraction in the range of $5-20\%$ will ensure even the smallest ionized region detectable by SKA-Low at $z \leq 10$ will contain tens of galaxies detectable above JWST and ELT spectroscopic thresholds. The JWST photometry will increase this number by more than a dozen; Roman and Euclid photometry will detect fewer galaxies than that. However, photometry comes with its own challenges, as the line-of-sight extent of the bubble ($\Delta z \approx 0.03 - 0.06$) is far smaller than the photometric redshift uncertainties ($\Delta z \sim 1$). Thus, photometrically detected galaxies within the bubbles suffer from confusion from interloper galaxies, whose number decreases with increasing redshift. Thus, a follow-up spectroscopic confirmation of the photometrically detected bubble galaxies via ELT-MOSAIC or JWST-NIRSpec would be essential for confirmation of their host bubble membership \citep{Chen+2025, Mason+2020, zackrisson_2020}.   

\section{The connection with the global 21-cm signal}

The global or sky-averaged 21-cm signal is another powerful observable. Although it does not carry any positional information, it captures the global evolution of the 21-cm signal and also measures its absolute strength. This section describes different ways in which SKA-Low measurements connect to the global signal. The first subsection describes a proposed method for measuring the global signal with the shortest baselines of SKA-Low and the second subsection outlines how high quality 21-cm image can be used to constrain the value of the global 21-cm signal.

\subsection{Measuring the global 21-cm signal with short baselines}

The global 21\,cm signal is a key observable for the thermal and ionization history during the Cosmic Dawn/EoR, yet it is obscured by bright diffuse foregrounds. Besides total-power experiments, very short baselines inside an SKA-Low station provide an interferometric response to a uniform sky, so that one station can be treated as a global-signal interferometer. 

For a two-element interferometer with baseline vector $\boldsymbol{b}$ at frequency $\nu$ (wavelength $\lambda=c/\nu$), the visibility is
\begin{equation}
V(\boldsymbol{b},\nu)=\frac{1}{4\pi}\int A(\hat{\boldsymbol{s}},\nu)\,T(\hat{\boldsymbol{s}},\nu)\,e^{-2\pi i(\boldsymbol{b}\cdot\hat{\boldsymbol{s}})/\lambda}\,{\rm d}\Omega,
\end{equation}
where $A$ is the primary beam pattern and $T$ is the sky temperature. Writing $T(\hat{\boldsymbol{s}},\nu)=T_0(\nu)+\delta T(\hat{\boldsymbol{s}},\nu)$ gives
\begin{equation}
V(\boldsymbol{b},\nu)=T_0(\nu)\,G(\boldsymbol{b},\nu)+V_{\rm ang}(\boldsymbol{b},\nu),\quad
G\equiv\frac{1}{4\pi}\int A\,e^{-2\pi i(\boldsymbol{b}\cdot\hat{\boldsymbol{s}})/\lambda}\,{\rm d}\Omega,
\end{equation}
where $G$ describes the interferometric response to a uniform sky. 
For short baselines with $\rho\equiv|\boldsymbol{b}|/\lambda\lesssim0.5$, the amplitude $|G|$ decreases smoothly as the baseline grows, reflecting the reduced sensitivity to the uniform sky component. This means that the coupling to the monopole remains strong on the shortest spacings, but rapidly weakens as the baseline length increases.

Prototype work has established the feasibility and practical requirements for these measurements. Using the Engineering Development Array-2 (EDA-2), the All-Sky SignAl Short-Spacing INterferometer (ASSASSIN) recovered a smooth Galactic spectrum from short spacings and emphasized the need for wide, smooth bandpasses, embedded-element beam/coupling models, and conservative baseline masks \citep{mckinley2020all}. The two-element Short-spacing Interferometer Telescope probing cosmic dAwn and epoch of ReionizAtion (SITARA) quantified correlated receiver noise and coupling on very short spacings and proposed calibration strategies \citep{thekkeppattu2022system,thekkeppattu2023singular}. The Short-spacing Interferometer for Global-signal Measurement and Analysis (SIGMA) concept is a purpose-built 1D short-spacing array that maintains $b<\lambda/2$ over 65–90\,MHz with explicit coupling treatment \citep{SIGMA_RAA}. Design and sensitivity considerations for interferometric monopole recovery are discussed in \citep{presley2015measuring,singh2015detection}.

This approach can be evaluated with a one-day single-station SKA Low simulation with $N=256$ elements. For this simulation drift scan visibilities are generated with OSKAR\footnote{\url{https://ska-telescope.gitlab.io/sim/oskar}} over 50–100\,MHz in 1\,MHz channels and 240\,s integrations. The antenna response is simulated in FEKO\footnote{\url{https://altairhyperworks.com/feko}} using the SKALA 4.1 design \citep{bolli2020test} to obtain the element beam patterns. Only cross correlations between distinct antenna elements are analyzed, applying a strict short baseline mask of $\rho\le0.5$.  Since the shortest physical spacing in the station is about 1.76\,m, no baseline satisfies the $\lambda/2$ condition above approximately 85.4\,MHz.

To extract the global sky component, a visibility-weighted estimator is used for each frequency channel using the short baseline data. Averaging these estimates over one sidereal day yields a smooth Galactic spectrum that follows the expected synchrotron power law between 50 and 80\,MHz. To test signal recovery, a flattened Gaussian profile similar to the EDGES measurement ($A=0.5$\,K) is injected at the visibility level using the unit sky kernel for Stokes~$I$. A two component fit at each $(\nu,t)$, $V \simeq a(\nu,t)\,G + b(\nu,t)\,M + n$, where $M$ is an angular foreground control derived from background only visibilities, followed by averaging of $a(\nu,t)$ over local sidereal time, successfully recovers the injected signal with an amplitude of $A_{\rm rec} = -0.505 \pm 0.002$\,K from the single day simulation. 

In extending this approach to real observations, the dominant limitations are expected to arise from instrumental chromaticity of the primary beam and bandpass, mutual coupling and correlated noise on the shortest spacings, polarization leakage that mixes foregrounds into Stokes~$I$, in-band reflections and slow gain variations, incomplete or imperfect sky models (including weak sidelobe sources), and environmental effects such as the ionosphere, radio interference, and near-field scattering. 
Future developments may explore strategies such as monitoring the spectral flatness of $G(\nu)$, performing visibility-level injection and recovery tests, applying conservative short-baseline selections, and combining data from multiple stations to reduce thermal noise and assess station-dependent systematics.

For band averaged amplitudes, the thermal uncertainty scales as
\begin{equation}
\sigma \propto \left(\Delta\nu\,\tau\,N_{\rm bl,eff}\,N_{\rm day}\right)^{-1/2}.
\end{equation}
In an idealized single-day simulation for SKA-Low, the thermal noise is already well below the level that would limit the analysis. 
Extending the integration to multiple days or combining measurements from several stations is therefore most valuable for improving internal consistency and allowing stricter short baseline selections, rather than for further reducing statistical noise. 
At the later stages of the SKA Array Assembly, such as AA4 and AA*, thermal sensitivity will no longer be the dominant constraint. 
The focus will instead shift toward controlling instrumental chromaticity and beam stability so that the effective coupling kernel $K(\nu)$ remains well characterized. 
Cross checks between stations, observing days, and baseline masks will then provide the basis for a credible detection of the global 21\,cm signal.

In summary, short baselines within an SKA-Low station enable direct interferometric estimation of the sky-averaged spectrum from visibilities, provided that the station kernel and bandpass remain smooth and stable. 
Our single-day simulation recovers both the smooth Galactic spectrum and the injected EDGES-like profile with high fidelity, in line with the experience from EDA2/ASSASSIN, SITARA, and SIGMA. 
These results point toward a clear, technically grounded path for achieving a credible detection of the global 21\,cm signal with SKA-Low.

\subsection{Measuring the global 21-cm signal from the images}

\citet{giri2018optimal} proposed a method to estimate the global signal by utilizing ionized regions identified in 21-cm signal images (see Section \ref{sec:structures_images} for methods to identify these ionized regions). Due to the absence of a zero-spacing baseline in interferometric observations, the 21-cm signal at each frequency channel has a mean value of zero. However, pixels within the ionized regions contain values that correspond to the negative of the global signal value at that frequency, and by inverting these pixel values, the global signal can be recovered.

This estimation method has important caveats, as it is sensitive to both foreground residuals and noise levels in the images. \citet{giri2018optimal} demonstrated that the method can be successfully applied to 21-cm signal images obtained with 1000-hour observations using the final SKA-Low configuration (AA4), provided that foreground residuals are subdominant and very large ionized regions are used for the measurement. Under these conditions, the technique can reliably recover the global signal from the 21-cm images.

\section{Conclusions}
SKA-Low will have the capability to produce tomographic image data of the redshifted 21-cm signal from the Cosmic Dawn and Epoch of Reionization. Producing high-fidelity image data will however be quite challenging and likely require substantial amounts of observing time ($> 1000$ hours) on selected fields and superb calibration and foreground removal techniques. As noise levels are lower at lower redshifts, the prospects are best for the reionization epoch, where the focus will most likely be on identifying ionized and neutral regions, effectively segmenting the image data into those two types of regions. Producing tomographic images data will be easier in AA4 than in AA$*$.

A wide range of analysis methods have been developed to quantitatively describe different aspects of such segmented image data, ranging from different algorithms extracting size distributions to many different types of topological quantities. Advances in machine learning are now also enabling analysis methods which do not rely on identifying specific features in the data. A different use of image data is to combine it in different ways with data from other telescopes for the same field.

Although interferometers in principle cannot measure the sky-averaged signal, techniques are being developed that allow extracting the global signal by using very short, intra-station, baselines, so essentially from very low resolution images.

\section{Author contributions}
The authors of this chapter contributed to its text in the following way.
SB, SD, AD, SM and SKP contributed to Sects.~2, 3.1, 4.1, 4.2. This includes the production of Fig.~3. SM also wrote the first version of Sect. 6.3.
MB contributed to the machine learning part of Section 3.1
SC, together with KKD and AM mainly contributed to Sect. 3.2 on the matched filtering technique, including Figs. 4 and 5.
IG wrote a large part of Sect. 4.1 on size distributions.
SKG primarily contributed to Sects. 2, 6.2 and 7.2 including producing Figure 1 and 2 and also helped with the text in Sect. 3.1.
IH wrote Section 5.1 on scattering transforms and their applications.
AK contributed to writing of Sect. 4.2 on topological quantities.
GM coordinated the writing of this chapter, wrote the abstract, Sects. 1, 8, the preambles of Sects. 3, 4, 5, 6 and 7, and was responsible for the final editing of the text. FZ contributed with the text of Sect. 7.1 on measuring the global signal with short baselines.

\bibliographystyle{abbrvnat-maxbibnames4}
\bibliography{chapter} 

\end{document}